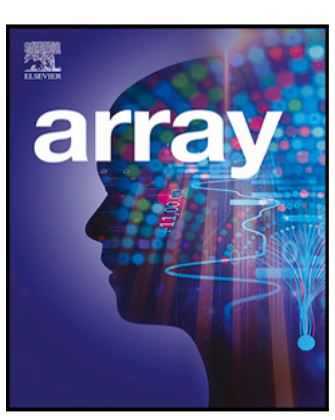

# Assessing projected quantum kernels for the classification of IoT data

Francesco D'Amore [a,*], Luca Mariani [a], Carlo Mastroianni [a], Francesco Plastina [b,c], Luca Salatino [a], Jacopo Settino [b,c], Andrea Vinci [a]

[a] Institute for high performance computing and networking (CNR-ICAR), Rende (CS), 87036, Italy
[b] Dipartimento di Fisica, Università della Calabria, Rende (CS), 87036, Italy
[c] INFN, Gruppo Collegato di Cosenza, Rende (CS), 87036, Italy

ABSTRACT

The use of quantum computing for machine learning is among the most promising applications of quantum technologies. Quantum models inspired by classical algorithms are developed to explore some possible advantages over classical approaches. A primary challenge in the development and testing of Quantum Machine Learning (QML) algorithms is the scarcity of datasets designed specifically for a quantum approach. Existing datasets, often borrowed from classical machine learning, need modifications to be compatible with current quantum hardware. In this work, we utilize a dataset generated by Internet-of-Things (IoT) devices in a format directly compatible with the proposed quantum data process, eliminating the need for feature reduction. Among quantum-inspired machine learning algorithms, the Projected Quantum Kernel (PQK) stands out for its elegant solution of projecting the data encoded in the Hilbert space into a classical space. For a prediction task concerning office room occupancy, we compare PQK with the standard Quantum Kernel (QK) and their classical counterparts to investigate how different feature maps affect the encoding of IoT data. Our findings show that the PQK demonstrates comparable effectiveness to classical methods when the proposed shallow circuit is used for quantum encoding.

## 1. Introduction

The use of quantum computing within machine learning frameworks has led to the emergence of *Quantum Machine Learning* (QML), a field that represents the convergence of two highly impactful disciplines and leverages the potential of both parent domains [1]. In [2], the authors highlight the methodological challenges of comparing classical and quantum machine learning, emphasizing that any claim of quantum advantage must be contextualized by clearly defining both the problem type and the nature of the data involved.

For the current NISQ (Noisy Intermediate-Scale Quantum) era, recent research has focused on the design, implementation and validation of variational hybrid algorithms. In these algorithms, quantum circuits, driven by tunable parameters, are adopted for tasks that can potentially bring a quantum speedup, while classical components are employed for the optimization of quantum circuit parameters and other tasks such as pre-processing and efficient storage of data [3–6]. Along this research avenue, our group has also reported results for several engineering optimization problems, such as energy management [7], resource allocation [8], image classification [9], and time series prediction [10].

Kernel methods are a cornerstone of classical machine learning, offering powerful tools for nonlinear data analysis through feature mapping and similarity measures. The Support Vector Machine (SVM) [11] is the most well-known kernel-based method. Although its use has declined with the rise of deep learning, kernel-based models remain relevant in quantum computing. In QML, kernel methods are particularly promising because: (i) quantum computers can map classical data into exponentially large Hilbert spaces, and (ii) the kernel between two data vectors corresponds to their state overlap, a natural quantum operation [12]. These features enable the exploration of richer data representations than those achievable with classical kernels. The *Projected Quantum Kernel* (PQK) approach [13] extends this concept by projecting quantum feature states back into a classical space through quantum measurements. This strategy mitigates the exponential dimensionality of the Hilbert space while preserving informative features extracted from quantum-encoded data.

In this work, we assess the flexibility and maturity of QML algorithms, and in particular Quantum Kernels, when processing real-world Internet of Things (IoT) sensor data. The task involves the classification of office room occupancy based on environmental parameters collected by indoor sensors. Our focus is on evaluating how effectively quantum models—particularly quantum kernel methods—can learn from realistic datasets, as well as on analyzing different design choices to identify the most effective ones.

* Corresponding author.
*E-mail address:* francesco.damore@icar.cnr.it (F. D'Amore).

Specifically, we compare PQK, the standard quantum kernel (QK), and their classical counterparts across multiple quantum encoding strategies (some introduced here) and measurement patterns. Our findings highlight the importance of encoding design and circuit depth in developing practical QML approaches for IoT data. Notably, shallow quantum circuits can match or even surpass deeper ones in terms of generalization, reinforcing trends previously observed in QML. As an example, in [14], the authors highlighted the effectiveness of shallow (low depth) and wide (more qubits) circuits for Variational Quantum Algorithms (VQAs). Shallow quantum classifiers could be aggregated into an ensemble model, as proposed for Variational Quantum Algorithms (VQAs) in [15]. This creates a more robust framework with effective error-mitigation capabilities.

Our experiments demonstrate that the performance of PQK is comparable to that of classical algorithms, although a definitive quantum advantage cannot yet be demonstrated. Furthermore, we observe that a moderate amount of shot noise—arising from a reduced number of circuit evaluations, or shots—can improve performance, in line with recent studies in other contexts [16]. We consider these findings promising and worthy of further investigation.

The paper is organized as follows: Section 2 introduces relevant background and state of the art regarding the use of artificial and real datasets, specifically generated in an IoT environment, as well as classical and quantum kernel approaches, in order to make the paper self-contained and accessible even to readers with limited knowledge of quantum computing; Section 3 describes the real-world dataset used for our study and the quantum circuits proposed and adopted for its encoding in a Hilbert quantum space; in Section 4, we describe the projected quantum kernel approach used for the experiments, as well as the sets of observables employed to map the quantum states onto the classical domain; Section 5 reports and discusses the main results achieved both with noiseless and noisy experiments; finally, Section 6 outlines potential avenues for future research in this domain.

This work extends our preliminary study presented in [17], providing a broader set of results, a more comprehensive overview for the reader, and a deeper exploration of the PQK. The developed code is publicly available on GitHub [18].

## 2. Background and related work

As anticipated in the introductory section, this section discusses the relevant background and state of the art in the main research areas addressed in this work.

### 2.1. Artificial and real-world datasets

To develop and assess machine learning models, researchers often rely on benchmark datasets, including the well-known and widely used MNIST. However, to process these datasets with current quantum hardware, a preprocessing step is typically required. This step involves downscaling or feature reduction techniques, often achieved through Principal Component Analysis (PCA) or using autoencoders (AE) [19].

As an example, in [20], authors propose the Projected Quantum Kernel Embedding based Link Prediction (PQKELP), a PQK approach on random walk embedding-based features to solve the link prediction problem. In order to test the proposed methods, they use five well-known datasets. In a data preprocessing step, the number of dimensions of the computing dataset is reduced using PCA, and then the reduced dataset is passed to a quantum circuit. Similarly, in [9], the authors use Quantum Extreme Learning Machines (QELM) for image classification on the MNIST dataset. To address the constraint of a limited number of qubits, they reduce the dimensionality of the data using PCA and AE.

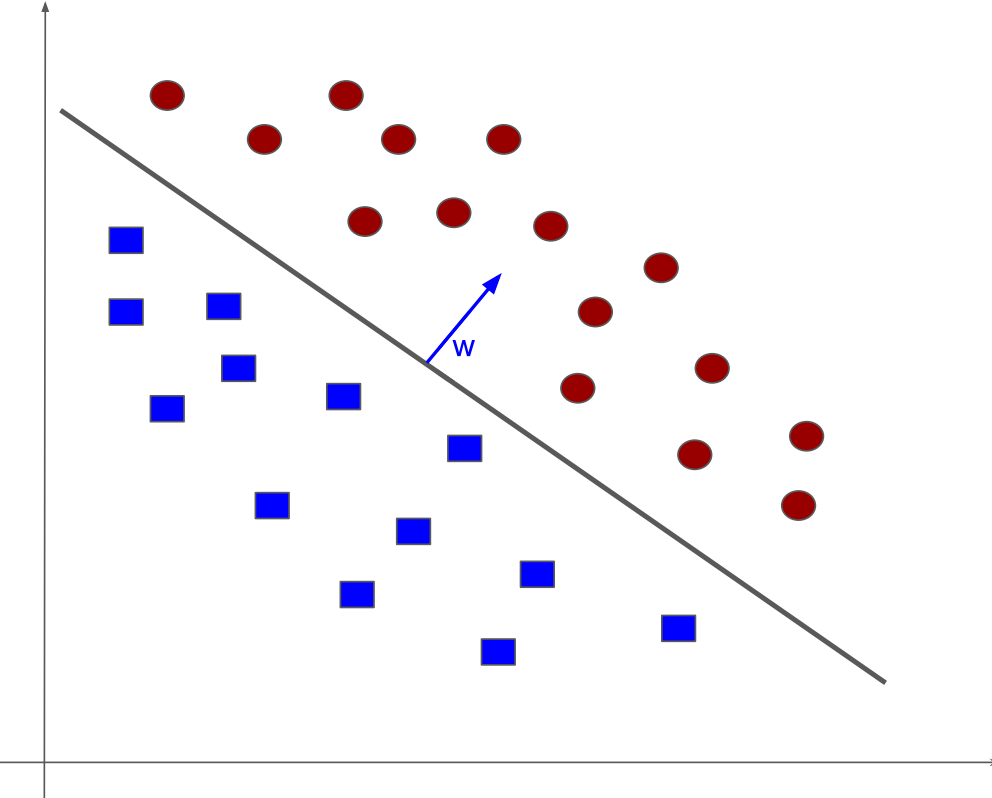

**Fig. 1.** In linear classification, the goal is to find a hyperplane that separates two classes. The slope of this hyperplane can be defined by a vector $\mathbf{w}$ orthogonal to it.

In [2], the authors highlight the potential limitations of using a benchmark dataset such as MNIST to compare traditional machine learning models with their quantum counterparts. Specifically, they argue that the necessary preprocessing steps can potentially distort the inherent nature of the problem under investigation. This can hinder meaningful comparisons with classical approaches or even with other quantum models that employ different feature reduction techniques. Furthermore, while techniques like AEs can generate robust data representations and enhance subsequent learning, they might obscure the unique advantages of quantum computing. One approach to overcome this issue is to use artificially generated data, which allows varying dataset properties and sizes [2,21,22]. However, with this approach, the conclusions may not easily generalize to real-world data. A more effective alternative is to use a real-world dataset that can be directly utilized without requiring further dimensionality reduction. For example, in [23] the authors used PQK to predict fermionic densities. The dimension of this physically informed dataset facilitates an agile comparison between classical and quantum kernel methods configured with various Hamiltonians, though the limited record count may constrain the generalization of the results.

In our study, we employed a real-world dataset that can be directly injected as input to a quantum computer, which facilitates a more direct comparison between different classical and quantum algorithms. The dataset will be discussed later, in Section 3.

### 2.2. Classical and quantum kernels

Kernel methods are a cornerstone of classical machine learning and are also increasingly important for their quantum counterpart. This is due to their elegant mathematical structure, which is heavily based on feature mapping and similarity metrics. The SVM [11] is the most well-known kernel-based method. For linearly separable problems, as the binary problem in Fig. 1, SVM can effectively find a hyperplane that separates the two classes. In Eq. (1) this hyperplane is defined by the weight vector $\mathbf{w}$ and bias $b$

$$g(\mathbf{x}) = \langle \mathbf{w}, \mathbf{x} \rangle + b \tag{1}$$

Finding the optimal hyperplane to separate two linearly separable classes can be formulated as a Quadratic Programming (QP) [24] problem, as shown in Eq. (2). In this equation, $y(\mathbf{x}) = \pm 1$ represents the ground truth label, i.e. the one associated to a given input vector $\mathbf{x}$ in the dataset.

$$\begin{aligned} \boldsymbol{\alpha} = \underset{\boldsymbol{\alpha}}{\operatorname{argmin}} \left( \frac{1}{2} \sum_{ij} y_i y_j \alpha_i \alpha_j \langle \mathbf{x_i}, \mathbf{x_j} \rangle - \sum_i \alpha_i \right) \\ \text{subject to: } \alpha_i \geq 0 \ \forall\, i, \quad \sum_i \alpha_i y_i = 0 \end{aligned} \tag{2}$$

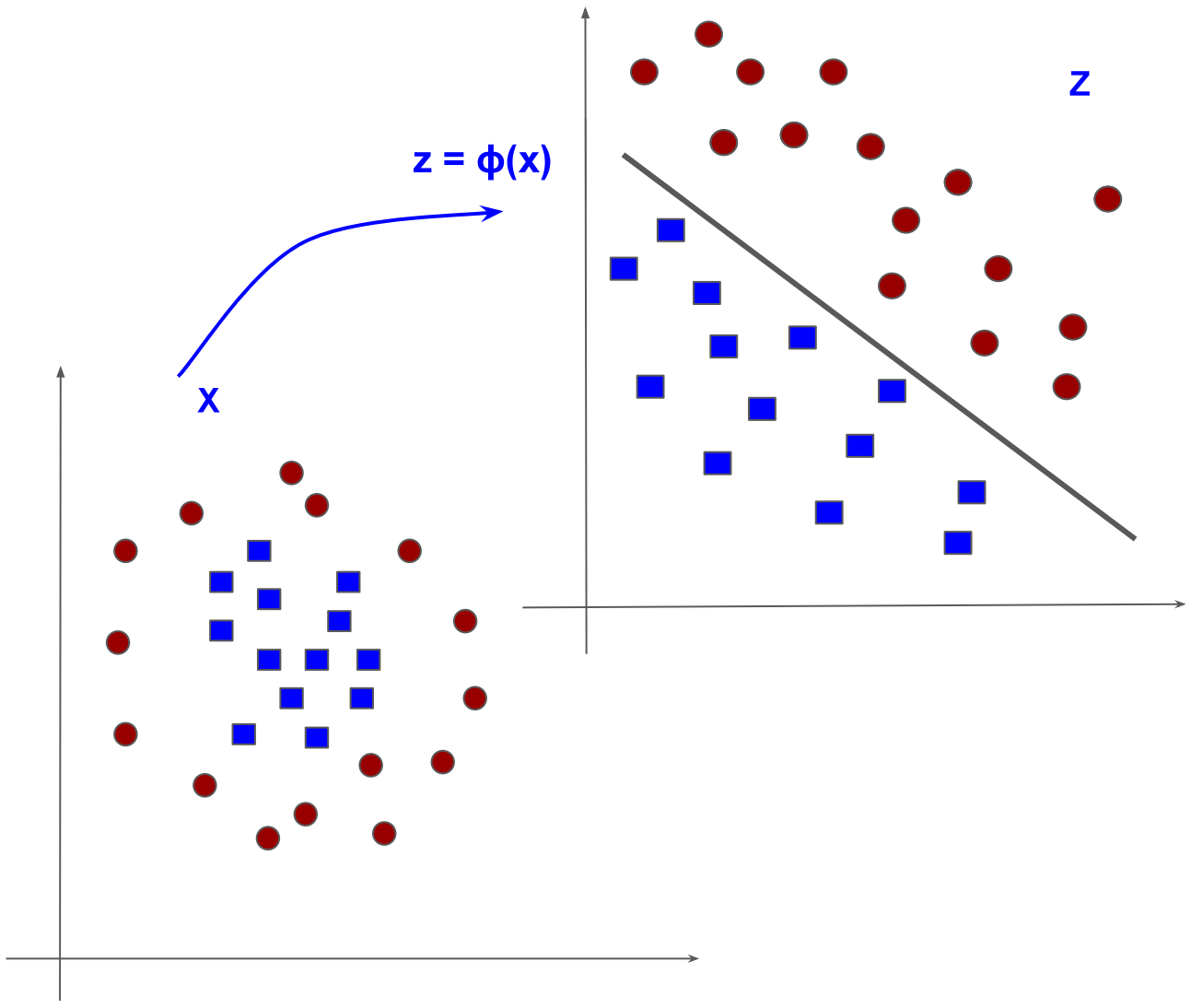

**Fig. 2.** Non-linear transformation can be employed to enhance the capabilities of linear classifiers by performing linear classification in the target space (Z).

The objective of Eq. (2) is to determine the Lagrange multipliers, $\boldsymbol{\alpha}$. A comprehensive discussion of QP in the context of SVM is beyond the scope of this work. For our purposes, we report the decision function in Eq. (3), which is derived from the results of Eq. (2).

$$\hat{y}(\mathbf{x}) = \text{sign}\left(\sum_i \alpha_i \langle \mathbf{x_i}, \mathbf{x} \rangle + b^*\right) \tag{3}$$

where $\hat{y}(\mathbf{x})$ denotes the label predicted by the model for the data-point vector $\mathbf{x}$, and the term $b^*$ depends on inner products of data points $\langle \mathbf{x_i}, \mathbf{x_j} \rangle$.

The length of the vector $\boldsymbol{\alpha}$ matches the number of data points in our dataset. Although Eq. (3) sums over all data points in the dataset, the vector $\boldsymbol{\alpha}$ is typically very sparse: non-zero values in this vector correspond to the eponymous "support vectors" $\mathbf{x_i}$, used to classify a new point $\mathbf{x}$. Due to the sparsity of the vector $\boldsymbol{\alpha}$, the number of support vectors is typically much lower than the total number of data points. Fewer support vectors, and therefore a more compact model, bring two advantages: (i) a reduced number of terms in the sum of Eq. (3), which improves the prediction speed on test data; (ii) better generalization properties, since the probability of committing a mistake on a test sample is bounded by the ratio between the expectation value of the number of support vectors and the number of training vectors [11].

For non-linearly separable data, as illustrated in Fig. 2, the previously described process can still be applied with minor modifications. In Fig. 2, data points from the $X$ space are mapped to the $Z$ space through a non-linear transformation $\mathbf{z} = \phi(\mathbf{x})$. In the Z space, transformed data points may become linearly separable by a hyperplane, which allows for the application of Eqs. (2) and (3). The kernel function itself is defined as the inner product in this $Z$ space, as shown in Eq. (4).

$$k(\mathbf{x}, \mathbf{x}') = \langle \phi(\mathbf{x}'), \phi(\mathbf{x}) \rangle \tag{4}$$

Using the transformation $\phi(\mathbf{x})$ and the kernel definition in Eq. (4), Eq. (3) becomes:

$$\begin{aligned} \hat{y}(\mathbf{x}) &= \text{sign}\left(\sum_i \alpha_i \langle \phi(\mathbf{x_i}), \phi(\mathbf{x}) \rangle + b^*\right) \\ &= \text{sign}\left(\sum_i \alpha_i k(\mathbf{x_i}, \mathbf{x}) + b^*\right) \end{aligned} \tag{5}$$

demonstrating that only the kernel function is required, and not the explicit form of the transformation. This embodies the core principle of the "kernel trick". To determine if a given function $k(\mathbf{x}', \mathbf{x})$ is a valid kernel, one must examine the Gram matrix (Eq. (6)), defined as the square matrix whose elements are the values of the kernel on all possible pairs of input vectors:

$$K_{ij} = k(\mathbf{x}_i, \mathbf{x}_j) \tag{6}$$

A function $k(\mathbf{x}', \mathbf{x})$ is a kernel if the corresponding Gram matrix $K_{ij}$ is positive semidefinite.

The optimal kernel for a classification task depends on the nature and distribution of the data points. One example is the Radial Basis Function (RBF), or Gaussian kernel, as defined below:

$$k_{\text{RBF}}(\mathbf{x}, \mathbf{x}') = \exp(-\gamma \|\mathbf{x} - \mathbf{x}'\|^2) \tag{7}$$

Its ability to capture complex, non-linear relationships between data points makes it a robust starting point for many classification problems. The factor $\gamma$ controls the behavior of the kernel in Eq. (7): it is the inverse of the standard deviation of the Gaussian distribution used within the kernel function. Greater values of $\gamma$ narrow the Gaussian curve, limiting the influence to nearby data points. A lower $\gamma$, on the other hand, widens the curve, allowing for a stronger interaction of the data points.

Although kernel methods have been widely employed, their use within the machine learning community has declined with the rise of other approaches, particularly neural networks and deep learning. In quantum computing, however, they are considered promising for two main reasons [12]: (i) quantum computers can map classical information into states that reside in exponentially large Hilbert spaces; (ii) the kernel between two data vectors can be computed through their overlap, a natural operation in quantum systems.

In classical kernel methods, the kernel trick offers the significant advantage of performing non-linear transformations without explicitly handling the function $\phi(\mathbf{x})$ that maps features to the new space. This stands in contrast to quantum computing, where classical data must be encoded into quantum space, necessitating explicit feature maps:

$$|\phi(\mathbf{x})\rangle = U(\mathbf{x})|0\rangle^{\otimes n} \tag{8}$$

The encoding defined in Eq. (8) can be considered an explicit quantum feature map that transforms classical data from the classical data space into the Hilbert space using the quantum circuit that applies the unitary $U(\mathbf{x})$. Developing an effective data encoding strategy and selecting the most suitable quantum circuit constitutes a significant area for future research, as the optimal feature map design may be highly dataset-dependent.

A quantum kernel can be thought of as the overlap of two quantum states [1]:

$$k(\mathbf{x}, \mathbf{x}') = |\langle \phi(\mathbf{x}') | \phi(\mathbf{x}) \rangle|^2 \tag{9}$$

In cases where the density operator is more convenient than the quantum state vector, Eq. (9) becomes:

$$k(\mathbf{x}, \mathbf{x}') = \text{Tr}\{\rho(\mathbf{x}')\rho(\mathbf{x})\} \tag{10}$$

where $\rho(\mathbf{x})$ is a data-dependent density matrix, similar to the data-dependent quantum state vector defined in Eq. (8). For pure states, Eq. (10) reduces to Eq. (9). There are several ways to compute the overlaps between quantum states, including the *Swap test* and the *adjoint method* [25], which combines Eqs. (9) and (8) to get:

$$k(\mathbf{x}, \mathbf{x}') = |\langle 0 | U^\dagger(\mathbf{x}) U(\mathbf{x}') | 0 \rangle|^2 \tag{11}$$

The majority of quantum kernels in the literature are computed via state overlaps, as described in Eqs. (9) and (10). Following this, the QP algorithm is applied, analogous to the classical SVM, to find the support vectors for the classification problem. An alternative to this approach is the Projected Quantum Kernel (PQK), which will be introduced in the next section.

**Fig. 3.** By projecting data points into quantum space, the dimensionality of the data representation is increased. This dimensionality is then mitigated by projecting back into a classical space, where the kernel trick can be exploited.

### 2.3. Projected quantum kernels

As discussed previously, quantum kernels rarely exploit the traditional kernel trick [26] because projection into the Hilbert space is performed explicitly, and the kernel is computed using only metrics such as the overlap between quantum states. An important exception is the Projected Quantum Kernel [13].

The encoding process is illustrated in Fig. 3. The unitary operator $U(\mathbf{x})$ maps the input data into the quantum space, where the Gram matrix is computed according to Eq. (11). While this approach leverages the high-dimensional representation of the Hilbert space to potentially enhance QML performance, this vast dimensionality can also hinder the learning process [22]. To mitigate this, the PQK approach addresses the exponential growth of the Hilbert space by employing a set of quantum measurements to project the quantum states back into a classical feature space. The resulting classical representations of the quantum states are used to calculate the elements of the Gram matrix $K_{ij}$ defined in Eq. (6). By introducing this intermediate quantum step, PQK can extract valuable features in the projected classical space.

This pattern is arguably analogous to the hidden layers of neural networks, where the intermediate layers and their associated data representations often have a higher dimensionality than the input and output layers. Similarly, as shown in Fig. 3, the quantum space is exponentially larger than the initial and final data space.

Using the PQK approach, the kernel and the relative Gram matrix can be evaluated using various methods, as outlined in [13]. Section 4 discusses the strategies adopted in this work.

## 3. IoT dataset and its quantum encoding

The primary objective of this study is to provide an overview of quantum kernels applied to a real-world IoT sensor dataset. Our case study allows for experimenting with different quantum kernel approaches and various data encoding circuits, ultimately enabling a comparison of the results.

### 3.1. The COGITO-IoT dataset

As anticipated in Section 2.1, the dataset adopted in this study can be directly injected as input to a quantum circuit, without any preprocessing, which facilitates a more direct comparison between different quantum models and their classical counterparts. This dataset, which contains 2865 records, has been previously employed in related research in the domain of machine learning for IoT [27], and it is publicly available. The data was acquired through COGITO [28], an IoT platform specifically designed for cognitive buildings. COGITO enables the implementation of holistic services that prioritize energy efficiency, security, and user comfort. For validation purposes, COGITO was deployed in an office building. Our goal is to predict room occupancy based on several environmental sensors regarding: illuminance, blind and lamp state, relative humidity (rh), carbon dioxide (co2), and temperature (temp). In particular:

- *illuminance* measures the intensity of light, which is likely high when the office is occupied;
- *blinds* indicates if windows are covered, potentially affecting illuminance;
- *lamps* is a direct indicator of artificial light usage, suggesting potential occupancy;
- *rh* (relative humidity) can be affected by human presence due to respiration;
- the CO2 (carbon dioxide) level rises with occupancy due to human respiration;
- *temp* (temperature) can be adjusted by occupants for comfort.

The 2865 observations were collected between November 1st and 25th, 2022, at the ICAR-CNR IoT Laboratory. This study focuses on data acquired during typical employee working hours. Fig. 4 illustrates the correlations among the features. The selected data first required filtering and cleaning to remove non-relevant information (such as $CO$ levels, which are irrelevant for human occupancy classification, and timestamps, which are useful for time-series prediction but not for classification), and redundant records. Subsequently, the processed data was normalized before being used in the quantum feature map described in Section 3.2 and in the classical SVM model.

The task is a binary classification problem: distinguishing between "occupied" and "non-occupied" states. The dataset exhibits minor class imbalance, with 56.5% (1619 samples) labeled as "non-occupied" and 43.5% (1246) labeled as "occupied".

### 3.2. Quantum encoding

Several techniques can be employed to map the input data $\mathbf{x}_i \in R^n$ to a quantum state $|\mathbf{x}_i\rangle$. The current state-of-the-art is generally oriented toward using a fixed, albeit often parameterized, ansatz and, in this work, we adhere to this strategy. We tested different circuits, starting from common options, and then moved toward shallower architectures based on the results gathered from initial attempts.

More specifically, we employed the three encodings proposed in [13], where the PQK was originally introduced. These encodings offer a good variety in terms of key characteristics (complexity, circuit depth, degree of entanglement), and their use enables a fair comparison with other studies on the same topic. Among these encodings, the simplest is defined by a qubit rotation circuit as in Eq. (12):

$$|\mathbf{x}_i\rangle = \bigotimes_{j=1}^{n} e^{-i\frac{x_{ij}}{2}X_j}|0^n\rangle, \tag{12}$$

where, $X_j$ is the Pauli-X operator acting on the $j$th qubit. This encoding will be referred to as "RotX". The remaining two circuits from [13] are more complex quantum encodings, specifically an Instantaneous Quantum Polynomial-time (IQP) encoding and a *Trotterized* circuit.

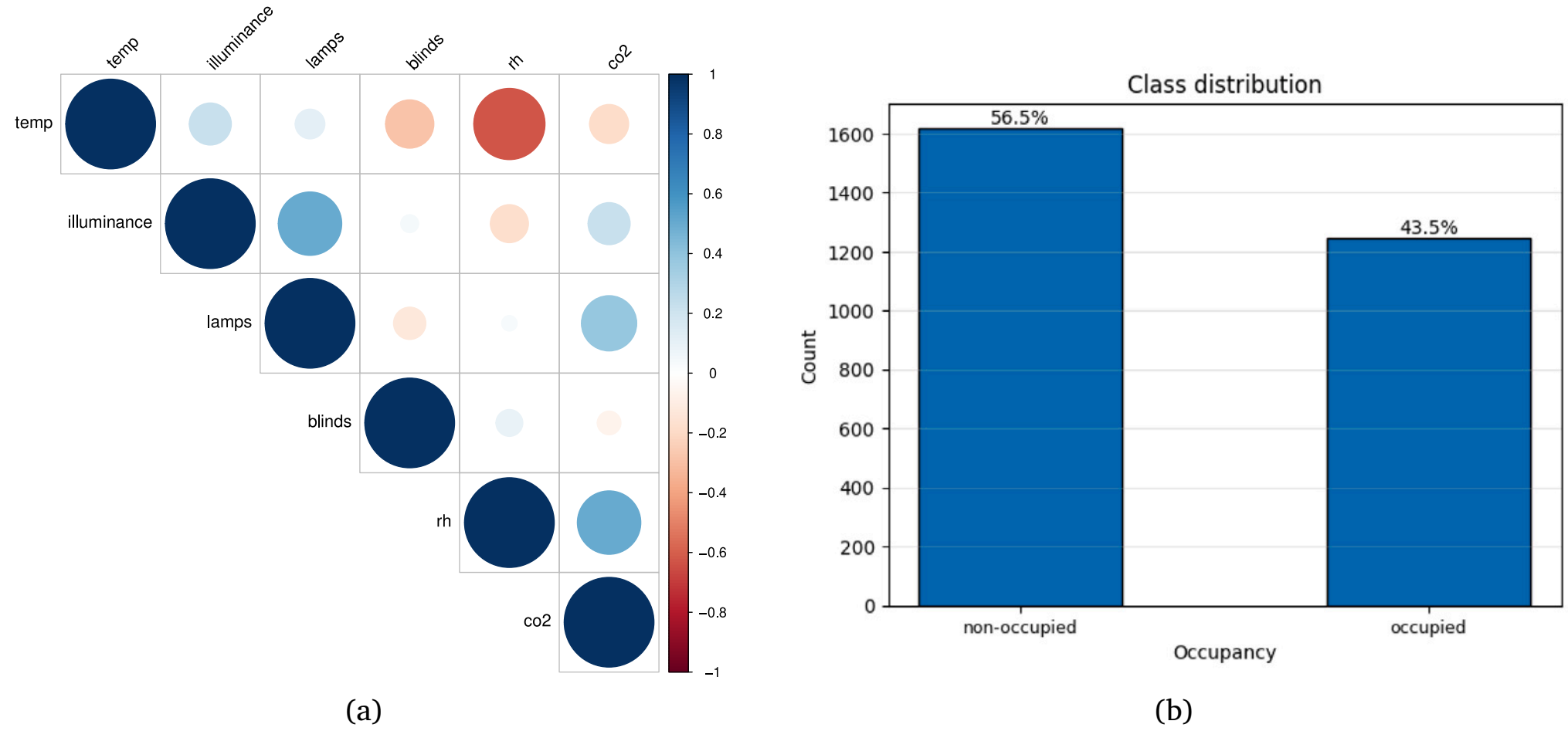

**Fig. 4.** (a) The correlation plot illustrates the relationships between features within the dataset. The color scale indicates both positive and negative correlations. (b) The class distribution plot illustrates the minor difference in distribution between the two labels present in the dataset.

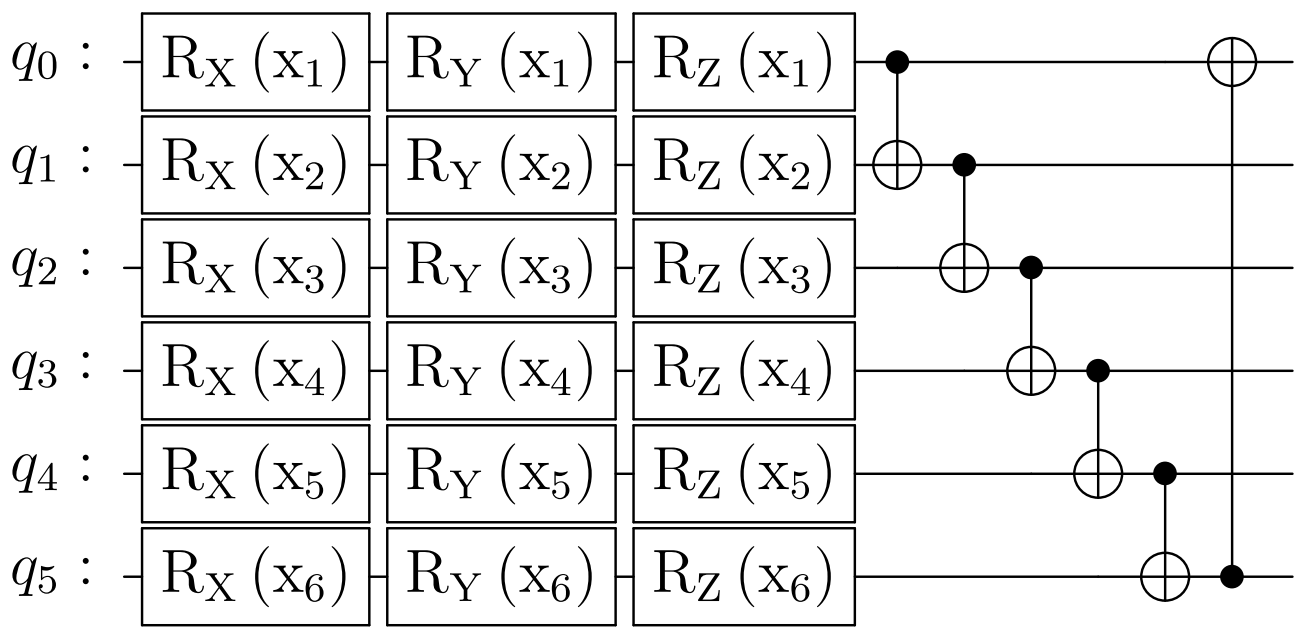

**Fig. 5.** The proposed quantum encoding is a variation of the encoding described in Eq. (12). This encoding rotates the qubits using the rotation gates on axes $X$, $Y$, and $Z$, in this order. We then investigate the role of entanglement by comparing the performances obtained with and without the controlled-NOT gates.

The *Trotterized* circuit requires specific insight. Following the approach in [13], we encode each classical datapoint $\mathbf{x}_i$ using Eq. (13), with $T = 20$ Trotter steps and an evolution time of $t = n/3$.

$$|x_i\rangle = \left( \prod_{j=1}^{n} \exp\left( -\frac{t}{T} x_{ij} (X_j X_{j+1} + Y_j Y_{j+1} + Z_j Z_{j+1}) \right) \right)^{T} \bigotimes_{j=1}^{n+1} |\psi_j\rangle, \tag{13}$$

The ZZFeatureMap is another common data encoding method. We tested it as an alternative strategy because it is widely used, including in similar application domains [29]. The main parameters governing the ZZFeatureMap are the number of circuit repetitions and the entanglement strategy. For this study, we adhered to the default configuration settings provided by the Qiskit implementation [30] (i.e., reps = 2, entanglement = "full").[1]

As later detailed in Section 5, experimental results suggest that shallow quantum encoding circuits outperform deeper ones; therefore, we designed and tested the circuit shown in Fig. 5 as a variant of the RotX circuit. This encoding uses six qubits, one for each feature of the dataset. Each qubit undergoes three rotations – one around each of the $X$, $Y$, and $Z$ axes, in this order – by an angle proportional to the corresponding feature value in the original dataset entry $\mathbf{x} = [x_1, x_2, x_3, x_4, x_5, x_6]$. Moreover, to evaluate the effect of entanglement on the performance of the resulting machine learning model, we tested the proposed quantum encoding with and without a set of circularly arranged controlled-NOT (CNOT) gates at the end of the circuit. In the following, the circuit in Fig. 5, without the CNOT gates, will be referred to as "3D", while the complete circuit including the CNOT gates will be referred to as "3D-CNOT". The name "3D" was chosen because the encoding involves three parameterized rotations around the $X$, $Y$, and $Z$ axes. All the circuits used in this work are reported in Fig. 6, with some simplifications as noted in the caption.

## 4. Projecting quantum states

A comprehensive overview of the data flow, from the initial dataset to the final machine learning model developed for this work, is provided in Fig. 7. The IoT dataset is transformed nonlinearly using quantum circuits, which both encode the data and perform a transformation similar to the one shown in Fig. 2. The purpose of the transformation is to map the data to a new space, enabling machine learning tools (SVM in this case) to analyze and classify them more effectively. Although the workflow depicts the quantum encoding proposed in Fig. 5, we compare the performances achieved with the different encoding circuits introduced above and reported in Fig. 6. Following the workflow, each data point $\mathbf{x}$ is mapped to a quantum state $|\phi(\mathbf{x})\rangle$ that serves as a representation of the original data in the Hilbert space. Following the PQK approach, as outlined in Section 2.3, a set of measurements is performed on the quantum state $|\phi(\mathbf{x})\rangle$. The outcomes serve as input for the calculation of the kernel, as mathematically defined by Eq. (14):

$$k_{\mathrm{PQK}}(\mathbf{x}, \mathbf{x}') = \exp\left( -\gamma \sum_{M \in S(n)_m} \sum_{O \in P} \left( \mathrm{Tr}\left( O\rho(\mathbf{x})_M \right) - \mathrm{Tr}\left( O\rho(\mathbf{x}')_M \right) \right)^2 \right) \tag{14}$$

[1] The adopted settings (in particular, the number of iterations in the Trotterized circuit and the number of repetitions in ZZFeatureMap) were motivated by two main considerations: (i) they represent a compromise between opposing requirements, and (ii) they are adopted as defaults in the literature; therefore, adhering to the same configurations enables a fair comparison with other studies, both those already published (e.g., [31,32]) and those that may wish to compare with our work in the future. For example, the parameter reps = 2 is often used as a practical choice, as it provides a good balance between circuit expressiveness (more repetitions can help to increase expressiveness) and overall depth (deeper circuits are more prone to noise and generally require longer convergence times). An analogous consideration applies to the choice of setting the number of Trotter steps to 20.

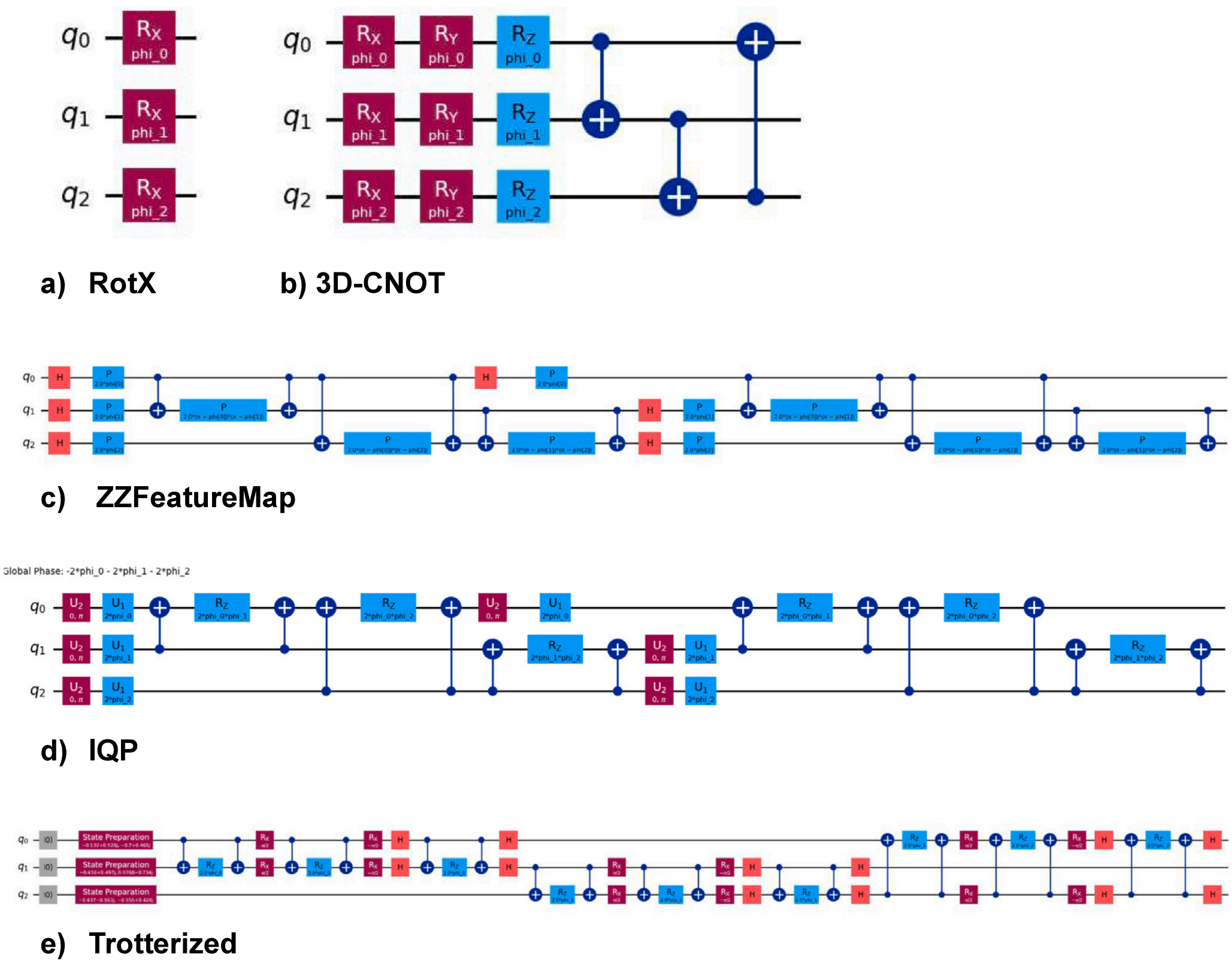

**Fig. 6.** This figure reports all the circuits used for quantum encoding. For representation purposes, the circuits are depicted with only three qubits, although our encoding uses six. Furthermore, the *Trotterized* circuit shown in *(e)* uses one Trotter step, whereas our experiments use 20 steps.

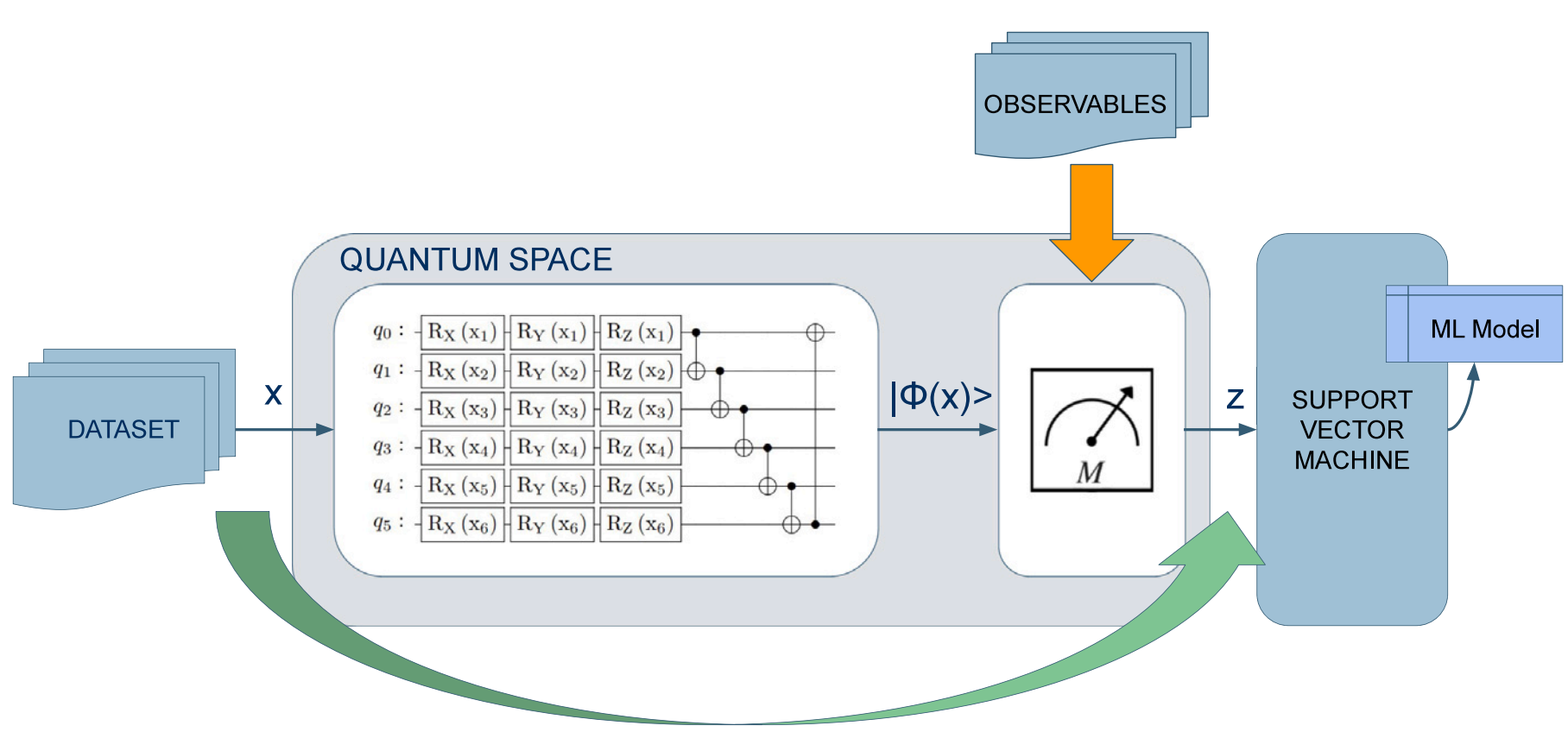

**Fig. 7.** The data pipeline. The dataset is mapped into the quantum space using a quantum circuit. Subsequently, the quantum states are projected back into a new classical space by performing measurements on selected observables. Notably, classical pre-processing of the data is not required using the proposed dataset, potentially leading to more meaningful comparisons between the performance of different ML models.

Here $M$ represents a subset of $m$ qubits selected among the $n$ qubits of the quantum circuit; $S(n)_m$ is the collection of admissible choices for $M$, and $\mathrm{Tr}\left(O\rho(\mathbf{x})_M\right)$ is the expected value of the observable $O$, extracted from a set of observables $\mathcal{P}$, and measured over the qubits in $M$. Here, the use of the Reduced Density Matrix (RDM) $\rho(\mathbf{x})_M$ in the formalism is essential to describe the state of the subsystem $M$.

There is considerable freedom in defining the projection strategies, which are determined by the choice of two sets: the set of observables, $\mathcal{P}$, and the set of qubit subsets, $S(n)_m$. In the context of this work, we considered a system with $n = 6$ qubits, since we associate a qubit with each feature of the dataset. Moreover, $\mathcal{P}$ always consists of Pauli operators, and the measurements are performed on single qubits ($m = 1$), on qubit pairs ($m = 2$), or on both:

- for $m = 1$ we take $M \in S(6)_1 = \{\{1\},\{2\},\ldots,\{6\}\}$: each feature in the classical space corresponds to the outcome of a local single-qubit measurement. Regarding the set of observables $\mathcal{P}$, at each qubit we measure the three Pauli operators $\mathcal{P} = \{X, Y, Z\}$, leading for each data point to a vector of 18 components in the classical feature space.
- for $m = 2$ we perform non-local measurements on the pairs of "adjacent" qubits; namely, we take $M \in S(6)_2 = \{\{1,2\},\{2,3\},\ldots,$ $\{6, 1\}\}$. $\mathcal{P}$ consists of tensor products of the three Pauli operators, specifically $\mathcal{P} = \{X \otimes X, Y \otimes Y, Z \otimes Z\}$. This strategy also yields data points represented as 18-component vectors within the classical space, facilitating meaningful comparisons among the projection methods.

- in the third scenario, the set of observables is defined as the union of the observables of the preceding two, with $m = 1$ and $m = 2$.

## 5. Experimental results

We performed a set of experiments to compare PQK with both non-projected Quantum Kernels (QK) and their classical counterparts (SVM), and assess the encoding circuits and projection strategies described above.

To perform our experiments, we provide a Python (v3.12.3) implementation of PQK using Qiskit (v1.3.1) [33] for quantum circuit management, and Pandas (v2.2.1) [34] together with scikit-learn (v1.4.2) [35] for data manipulation and basic SVM analysis. Our library allows users to specify both the quantum encodings and the observables used to project the quantum states to the classical space. Qiskit Machine Learning (v0.7.2) [36] was employed to simulate non-projected Quantum Kernels. The code, along with the experimental results discussed in this work, is available on GitHub [18]. For the first set of experiments, we used Qiskit's noiseless Statevector simulator, which maintains the complete representation of the quantum state. Measurements obtained with this simulator are exact and correspond to the values that would be achieved with an infinite number of circuit evaluations, or *shots*.

Table 1 reports the accuracy values for room occupancy prediction, along with the F1-score and the number of support vectors (#SV), for all the configurations discussed in the previous section, including classical SVM. In particular, we used all the quantum encoding circuits discussed in Section 3.2 to assess PQK and QK. For PQK in particular, we also considered different strategies for projecting quantum states (Section 4).

For the classical SVM, we used the RBF kernel as defined in Eq. (7), while for PQK we adopted the kernel defined in Eq. (14). In both cases, the following two key hyperparameters were tuned: the factor $\gamma$ in Eqs. (7) and (14) and the regularization parameter $C$, which controls the trade-off between the training error and the width of the margin [24]. In particular, setting a large value for $C$ forces the algorithm to construct a hard margin hyperplane that perfectly classifies the training data. Decreasing the value of $C$ creates instead a soft margin by allowing for some misclassifications in the training set, which can potentially improve the model's generalization to unseen data. In the case of non-projected Quantum Kernels, we tuned only the parameter $C$ because the $\gamma$ factor is not required.

The Gram matrix represents a pivotal intermediate stage across all evaluated approaches. Since our quantum feature maps utilize Pauli measurements bounded within $[-1, 1]$, the SVM bias term $b$ (Eq. (3)) is arguably sufficient to manage feature space offsets. Consequently, explicit centering of the Gram matrix is omitted, as it is often redundant in standard SVM implementations where the optimization naturally compensates for such shifts [37].

Table 1 reports the average accuracy and $F1$-scores, as well as the corresponding confidence intervals under the best hyperparameters. The optimal values of $C$ and $\gamma$, defined as the ones that lead to the highest score, were determined through *grid search* over the sets:

$$C \in \{0.006, 0.015, 0.03, 0.0625, 0.125, 0.25, 0.5, 1.0, 2.0, 4.0, 8.0, 16.0, 32.0, 64.0, 128.0, 256, 512, 1024\}$$

$$\gamma \in \{0.03, 0.0625, 0.125, 0.25, 0.5, 1.0, 2.0, 4.0, 5.0, 7.0, 8.0\}$$

In order to select the best option, a 10-fold cross-validation was used to estimate the average scores and corresponding standard deviations for each pair of parameter values within the specified grid. Subsequently, 95% confidence intervals were computed as twice (2x) the Standard Error (SE), defined as $SE = \sigma/\sqrt{n}$, where $\sigma$ is the standard deviation and $n$ is the number of samples. In our case, the number of samples ($n$) is 10, i.e., the number of values produced by the 10-fold cross-validation procedure. To ensure the reproducibility of the cross-validation process, we fixed the random seeds for our experiments. We also verified the robustness of our findings by varying these seeds across multiple runs; the results remained qualitatively consistent with those reported herein.

**Table 1**
The table reports the accuracy, number of support vectors (#SV), and F1-score for models developed using various quantum encoding and measurement strategies. The last row shows the classic SVM. Scores that are competitive with the classical model are bolded. Confidence intervals were estimated as ±2 times the standard error of the mean (SE).

| Encodings | | Accuracies and #SV | | F1-Score |
|---|---|---|---|---|
| Kernel | Ft. map | $\mathcal{P}$ = {X,Y,Z} | | |
| $m = 1$ PQK 18 obs | RotX | **0.894 ± 0.008** | 1025 | **0.885 ± 0.008** |
| | 3D-CNOT | 0.890 ± 0.007 | 890 | 0.880 ± 0.007 |
| | 3D | **0.894 ± 0.008** | 970 | **0.883 ± 0.009** |
| | ZZFeatureMap | 0.750 ± 0.021 | 2578 | 0.697 ± 0.027 |
| | IQP | 0.879 ± 0.009 | 1445 | 0.863 ± 0.010 |
| | Trotterized | 0.873 ± 0.010 | 1720 | 0.857 ± 0.012 |
| Kernel | Ft. map | $\mathcal{P}$ = {XX,YY,ZZ} | | |
| $m = 2$ PQK 18 obs | RotX | 0.889 ± 0.008 | 1047 | 0.878 ± 0.009 |
| | 3D-CNOT | **0.901 ± 0.008** | 1085 | **0.892 ± 0.008** |
| | 3D | **0.895 ± 0.009** | 1062 | **0.884 ± 0.010** |
| | ZZFeatureMap | 0.730 ± 0.012 | 1997 | 0.680 ± 0.016 |
| | IQP | 0.865 ± 0.010 | 1437 | 0.849 ± 0.008 |
| | Trotterized | 0.886 ± 0.008 | 1217 | 0.872 ± 0.009 |
| Kernel | Ft. map | $\mathcal{P}$ = {X,Y,Z,XX,YY,ZZ} | | |
| $m = \{1, 2\}$ PQK 36 obs | RotX | 0.892 ± 0.008 | 1005 | 0.881 ± 0.009 |
| | 3D-CNOT | **0.897 ± 0.007** | 1127 | **0.887 ± 0.007** |
| | 3D | **0.895 ± 0.009** | 824 | **0.884 ± 0.010** |
| | ZZFeatureMap | 0.783 ± 0.011 | 2482 | 0.736 ± 0.011 |
| | IQP | 0.889 ± 0.009 | 1184 | 0.876 ± 0.010 |
| | Trotterized | 0.889 ± 0.012 | 1176 | 0.875 ± 0.014 |
| Kernel | Ft. map | Accuracies and #SV | | F1-Score |
| QK | RotX | 0.865 ± 0.008 | 872 | 0.855 ± 0.010 |
| | 3D-CNOT | 0.881 ± 0.012 | 836 | 0.870 ± 0.012 |
| | 3D | 0.881 ± 0.012 | 836 | 0.870 ± 0.012 |
| | ZZFeatureMap | 0.806 ± 0.013 | 2000 | 0.775 ± 0.015 |
| | IQP | 0.884 ± 0.004 | 1147 | 0.870 ± 0.005 |
| | Trotterized | 0.889 ± 0.010 | 939 | 0.871 ± 0.011 |
| SVM | RBF | 0.893 ± 0.009 | 1014 | 0.883 ± 0.009 |

In our opinion, the most interesting outcomes are the following:

- In terms of accuracy and F1-score, PQK achieves comparable or superior performance to QK when using shallower feature maps, i.e., RotX, 3D-CNOT, and 3D. Notably, a paired t-test reveals that the best PQK model significantly outperforms the best QK model (Trotterized feature map). The difference $\delta$ observed between the scores of the two models is modest but statistically significant. This is supported by the $p$-values ($p < 0.01$ for both accuracy and F1-score) and the corresponding 95% confidence intervals: $\delta_{acc} \in [0.0055, 0.0182]$ and $\delta_{F1} \in [0.0136, 0.0282]$, indicating a non-zero and consistent performance gap. The corresponding effect size, measured by Cohen's d, exceeds 1.3, indicating a large and practically meaningful difference between the two models. However, this advantage does not occur when using deeper feature maps. This suggests that, for the data transformations required in machine learning, single-qubit rotations may be more effective than complex encodings like the ZZFeatureMap, which involves intricate entanglement strategies. QML developers should carefully consider the data encoding, as deeper and more complex circuits do not appear to guarantee improved performance in learning models. Entanglement appears to have limited utility in creating effective machine learning prediction models, in our case. In fact, this type of outcome has been explored more comprehensively in previous studies [2].

- PQK can match the performance of the classical SVM-RBF. Among the tested models, the best-performing is the one that encodes the data through the 3D-CNOT feature map shown in Fig. 5 and performs the set of measurements with $m = 2$. More specifically, PQK and SVM-RBF yield very similar results across all feature maps except ZZFeatureMap and, in some cases, PQK even outperforms SVM-RBF in terms of accuracy and F1-score (see the highlighted rows in Table 1). In particular, the best PQK model exhibits an improvement in F1-score compared to the SVM-RBF, which is supported by the paired t-test ($p = 0.024$), by the 95% confidence interval ($\delta_{F1} \in [0.0013, 0.0147]$) and by a large effect size (Cohen's $d = 0.85$). While this study suggests that PQK can become competitive with the classical SVM, it shows once again that carefully choosing the quantum encoding is essential, since a deeper and more complex encoding does not necessarily guarantee better performance.
- PQK provides more compact models with respect to the classical SVM-RBF approach. Indeed, when using the 3D feature map and the set of observables $\mathcal{P} = \{X, Y, X, XX, YY, ZZ\}$, PQK allows for a 18% reduction in the number of support vectors when compared to SVM-RBF. A reduction is also achieved with QK with the three shallower feature maps. Compact models, which offer the advantages discussed in Section 2.2, suggest that both projected and non-projected quantum kernels could perform well with less data than their classical counterparts.

### 5.1. Scalability and computational considerations

It is interesting to assess the cost, in terms of computing time, of the kernel algorithms. The cost is evaluated in terms of the number of records $N$ and the number of features $n$, which in our scenario is equal to the number of qubits when using the quantum algorithms. It is convenient to partition the evaluation into two components; (*i*) the quantum component, which is present only in the QK and PQK algorithms, and (*ii*) the classical component, which is present both in the quantum algorithms and in the classical SVM (see Fig. 7).

The quantum component, adopted only for the QK and PQK algorithms, requires the evaluation of the cost related to the feature map. This cost, $C_{Fmap}$, is a function of the depth of the quantum circuit ($D$), multiplied by the number of commuting-observable measurement groups ($J$) and by the required number of measurement shots ($S$), yielding $C_{Fmap} \approx \mathcal{O}(D \cdot J \cdot S)$. Regarding the depth $D$, we consider the circuit depth in terms of two-qubit gates only, since they represent the bottleneck of current quantum hardware. Therefore, $D$ is a function of $n$, which depends on the type of circuit. With a simple analysis, it can be seen that $D$ is $\mathcal{O}(1)$ with the feature maps RotX and 3D, $\mathcal{O}(n)$ for the feature maps 3D-CNOT and Trotterized, and $\mathcal{O}(n^2)$ with the feature maps ZZFeatureMap and IQP. The number of measurements is different in the QK and PQK cases. With QK, the qubit measurements are performed only on the computational basis ($Z$), therefore $J = 1$. In our implementation of PQK, we need to measure observables on all three axes, $X$, $Y$, and $Z$, thus $J = 3$. The component $J$ is thus negligible in an asymptotic analysis, as it can be treated as a constant. The factor $S$, i.e., the number of shots, is specifically discussed in the next section. Finally, it must be considered that each feature map must be executed $N^2$ times in QK (because the overlap must be computed for each couple of records), and $N$ times in PQK (because each record needs to be projected into the classical space through the measurement operations).

When analyzing the classical component, it should be considered that both the Quantum Kernels and their classical counterparts rely on solving the QP problem introduced in Eq. (2) to find the optimal Lagrange multipliers. To do this, we consider the cost of computing the Gram matrix in the classical component for PQK and classical SVM, which corresponds to a cost of $\mathcal{O}(N^2 \cdot n)$, while, for QK, the Gram matrix is provided by the quantum component. Using algorithms like Sequential Minimal Optimization (SMO) [38] with a pre-computed Gram matrix, the practical training time of the SVM is $\mathcal{O}(N^2)$. This results, for SVM and PQK, in a time complexity of $\mathcal{O}(N^2 \cdot n)$, which includes both the computation of the Gram matrix and the solution of the QP problem in Eq. (2). Instead, for QK, the classic time complexity is $\mathcal{O}(N^2)$, since the Gram matrix has been computed in the quantum component. Since, in general, $n \ll N$, we can state that the time complexity of the classical components is $\mathcal{O}(N^2)$ for all the algorithms.

In conclusion, for PQK, since the computing time of the quantum part scales linearly with $N$, the overall computing time is asymptotically dominated by the classical part. Conversely, for QK, both the quantum and the classical sections scale with $\mathcal{O}(N^2)$. Therefore, the PQK approach has the benefit of reducing the quantum computing time to $\mathcal{O}(N)$; thus, providing an advantage over QK in terms of overall complexity, while also yielding better quantitative metrics as discussed before.

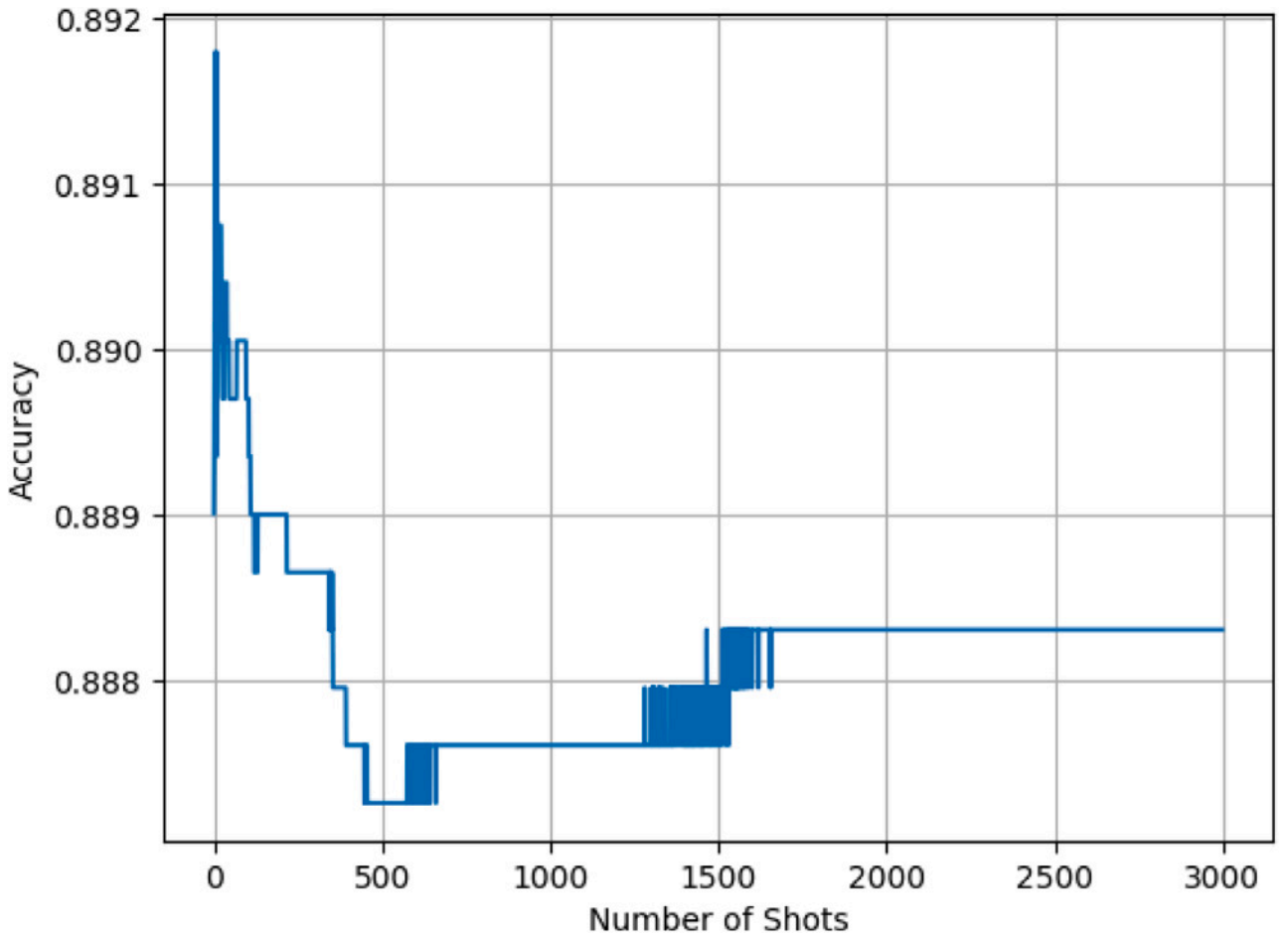

**Fig. 8.** Accuracy vs. the number of shots, with $m = 2$ and the 3D-CNOT feature map.

### 5.2. Adding shot noise

The evaluation of noise in quantum computation is a task with several different facets, some of which are quite peculiar when compared to classical computation. One of such aspects is that the expected values of the observables are calculated on a finite number of shots, which introduces intrinsic *finite-sampling noise*, or *shot noise*, even in ideal quantum computers. It is known that shot noise decreases as $O\left(1/\sqrt{N_{\text{shots}}}\right)$ as the number of shots $N_{shots}$ increases [39]. The impact of shot noise and techniques for its mitigation have been the focus of recent studies [40,41]. In line with this, after evaluating the models using exact quantum measurements (as reported in Table 1), we investigated instead how using a finite number of shots affects the performances. Specifically, we varied the number of shots to explore how the precision of measurement influences the accuracy of the model. For this study, we selected, among the configurations of Table 1, the one with $m = 2$ and the shallow 3D-CNOT feature map, previously shown in Fig. 5. The results are reported in Fig. 8.

Since each shot yields a distinct classical representation of the quantum states, a 10-fold cross-validation was employed for each shot count to ensure a more robust performance evaluation. However, we kept the values of $C$ and $\gamma$ constant for all shot counts because we are not interested in reproducing the accuracy value of Table 1, but rather to investigate the impact of the inherent stochasticity of quantum measurements on the QML model.

Fig. 8 shows that the accuracy appears to converge to a final value with a relatively small number of shots, i.e., about 1500. This suggests that further enhancement of the quantum measurement precision, beyond a relatively low threshold, does not provide any appreciable benefit. This aligns with the widely accepted principle in NISQ computing that shallower quantum circuits are inherently less vulnerable to noise accumulation than deeper ones, making their adoption preferable even when considering noise effects.

Moreover, experiments conducted with the introduction of shot noise revealed that shallow circuits achieve the highest accuracy with a small number of shots, typically ranging from a few dozen to around 300. This suggests that, unlike in other areas of quantum computing, where noise is regarded solely as detrimental, shot noise in QML algorithms can, in fact, exert a positive influence on model performance.

This observation is consistent with recent studies on the beneficial role of noise in VQA optimization. Specifically, a very recent study showed that shot noise can facilitate escape from saddle points and promote improved convergence [16]. Similarly, in the protocol proposed in [42], the authors use a targeted quantum noise injection to smooth and regularize quantum loss landscapes, thereby alleviating the problem of poor local minima.

## 6. Conclusions and future directions

The lessons learned from testing quantum circuits in a machine learning task suggest that significant challenges remain to effectively integrate quantum computing as a valuable tool for machine learning and artificial intelligence. By testing QML on real-world datasets, as done in this work, we can effectively assess the adaptability of these methods. The results demonstrate that PQKs are an effective alternative to their classical counterparts. Moreover, PQKs provide a moderate but statistically significant advantage over QKs in terms of quantitative metrics (accuracy and F1 scores), while also reducing the overall complexity of the quantum computation. Although this work confirms that a clear advantage has not yet been demonstrated, there are intriguing aspects that warrant further investigation.

In a generic machine learning pipeline, such as the one illustrated in Fig. 7, one key novelty introduced by the field of Quantum Machine Learning lies in the quantum encoding stage. Although encoding classical data into circuits is essential in any quantum computation, in QML it plays a crucial role [1]. The results of this work further support this notion. Empirical results obtained from a real-world dataset suggest that, in both QK and PQK contexts, shallow circuits can outperform deeper ones. While this outcome is expected in noisy circuits — since noise effects grow with circuit depth — it is noteworthy that the same holds in noiseless experiments.

From an architectural perspective, it is worthwhile to compare quantum kernel models with their classical counterparts, focusing on hyperparameters and model structure. All parameters not directly involved in the learning process should be regarded as hyperparameters. As previously discussed, hyperparameter tuning in SVMs typically involves selecting a kernel function, the regularization parameter $C$, and, when applicable, the $\gamma$ factor. In contrast, quantum kernels require a circuit to encode classical data into quantum states, while PQK models additionally demand a strategy to project the quantum state back into the classical domain. Since these elements are not directly part of the learning process, they should be considered additional hyperparameters. These design choices require careful consideration when developing a QML pipeline: although a larger number of hyperparameters can increase model flexibility and potentially improve performance, it also entails higher computational costs and more demanding tuning efforts.

Based on the results of our experiments, we deem that it is important to investigate the following points, also in other scenarios: (i) confirm that deep circuits are not always necessary to achieve higher model performance, which may serve as a best practice for researchers and developers; (ii) compare QML models trained with high-precision quantum measurements (large numbers of shots) against those using lower-precision measurements with a moderate level of shot noise (fewer shots); (iii) assess the potential positive impact on model performance resulting from the introduction of shot noise.

Furthermore, while the selected measurement strategies significantly influence the behavior of a PQK, a dynamic or adaptive measurement strategy could be further designed to adapt to the learning task and thereby improve data representation in the projected classical space. This approach could represent a particularly compelling research avenue to explore when combined with a Trainable Quantum Kernel [43], where the quantum feature map utilizes variational parameters that can be classically tuned to enhance the overall model performance.

It is important to note that the findings of this study are based on a single dataset collected from a specific office environment. While evaluating QML procedures in real-world scenarios is highly valuable, future work should extend this analysis to larger, more diverse datasets from different locations or application domains. Such efforts will be crucial to generalize our findings on the performance of shallow circuits and to evaluate further how projection design influences overall model performance.

Finally, our current analysis employs a noiseless simulator, which is adequate for initial tests on shallow circuits and shot noise. The next step involves moving to real noisy hardware. A hardware implementation would significantly improve our understanding of how circuit depth impacts Quantum Kernel Machines and how error sources must be addressed when developing machine learning models.

## CRediT authorship contribution statement

**Francesco D'Amore:** Writing – review & editing, Writing – original draft, Visualization, Validation, Supervision, Software, Methodology, Investigation, Formal analysis, Data curation, Conceptualization. **Luca Mariani:** Writing – review & editing, Writing – original draft, Validation, Software, Methodology, Formal analysis, Data curation, Conceptualization. **Carlo Mastroianni:** Writing – review & editing, Writing – original draft, Supervision, Project administration, Methodology, Investigation, Funding acquisition, Formal analysis, Data curation, Conceptualization. **Francesco Plastina:** Writing – review & editing, Supervision, Methodology, Investigation, Formal analysis, Conceptualization. **Luca Salatino:** Writing – review & editing, Writing – original draft, Visualization, Methodology, Investigation, Formal analysis. **Jacopo Settino:** Writing – review & editing, Writing – original draft, Validation, Methodology, Formal analysis, Conceptualization. **Andrea Vinci:** Writing – review & editing, Writing – original draft, Validation, Software, Methodology, Investigation, Funding acquisition, Data curation, Conceptualization.

## Declaration of Generative AI and AI-assisted technologies in the writing process

During the preparation of this work the author(s) used Gemini and Writefull in order to improve language and readability of the paper. After using this tool/service, the author(s) reviewed and edited the content as needed and take(s) full responsibility for the content of the publication.

## Declarations and funding support

This work was partially funded by

- ICSC – Italian Research Center on High Performance Computing, Big Data and Quantum Computing, funded by European Union – NextGenerationEU, PUN: B93C22000620006.

- "INSIDER: INtelligent ServIce Deployment for advanced cloud-Edge integRation" granted by the Italian Ministry of University and Research (MUR) within the PRIN 2022 program and European Union - Next Generation EU (grant n. 2022WWSCRR, CUP H53D23003670006, B53D23013250006).
- PNRR MUR project PE0000023-NQSTI through the secondary projects "QuCADD" and "ThAnQ".
- European Union – NextGenerationEU – National Recovery and Resilience Plan (Piano Nazionale di Ripresa e Resilienza, PNRR) – Project: "SoBigData.it – Strengthening the Italian RI for Social Mining and Big Data Analytics" – Prot. IR0000013 – Avviso n. 3264 del 28/12/2021.

## Declaration of competing interest

The authors declare that they have no known competing financial interests or personal relationships that could have appeared to influence the work reported in this paper.

## Data availability

The dataset required to reproduce the findings of this study is openly available in the GitHub repository: https://github.com/fdamore/QK/blob/main/origin/env.sel3.csv. This repository also contains the source code used for data processing and analysis (D'Amore et al., 2025)